\documentclass{aipproc}

\usepackage{graphicx} 
\usepackage{amssymb}

\layoutstyle{6x9}

\begin{document}
\title{Entropy considerations in constraining the mSUGRA parameter space
\footnote{To appear in the {\bf Proceedings of X Mexican Workshop on Particles 
and Fields}, Morelia Michoac\'an, M\'exico, November 7-12, 2005.}}

\classification{}
\keywords{dark matter, supersymmetry}

\author{Dario Nunez, Roberto A. Sussman, Jesus Zavala, Lukas Nellen}
{address={Instituto de Ciencias Nucleares,\\
             Departamento de Gravitaci\'on y Teor\'{i}a de Campos,\\
             Universidad Nacional Aut\'onoma de M\'exico (ICN-UNAM).\\
             A. Postal 70-543, 04510 M\'exico, D.F., M\'exico.}}
\author{Luis G. Cabral-Rosetti}
{address={Departamento de Posgrado,\\
Centro Interdisciplinario de Investigaci\'on y 
Docencia en Educaci\'on T\'ecnica (CIIDET),\\
Av. Universidad 282 Pte., Col. Centro, A. Postal 752, C. P. 76000,\\
Santiago de Quer\'etaro, Qro., M\'exico.}}
\author{Myriam Mondrag\'on}{address={Instituto de F\'{\i}sica,\\
             Universidad Nacional Aut\'onoma de M\'exico (IF-UNAM).\\
             A. Postal 20-364, 01000 M\'exico D.F., M\'exico.}}
\begin{abstract}
\noindent
We explore the use of two criteria to constraint the allowed parameter space in 
mSUGRA models. Both criteria are based in the calculation of the present 
density of neutralinos as dark matter in the Universe. The first one is the 
usual ``abundance'' criterion which is used to calculate the relic density 
after the ``freeze-out'' era. To compute the relic density we used the 
numerical public code micrOMEGAs. The second criterion applies the 
microcanonical definition of entropy to a weakly interacting and self-gravitating gas 
evaluating then the change in the entropy per particle of this gas between the 
``freeze-out'' era and present day virialized structures. An 
``entropy-consistency'' criterion emerges by comparing theoretical and empirical
estimates of this entropy. The main objective of our work is to determine for
which regions of the parameter space in the mSUGRA model are both criteria 
consistent with the 2$\sigma$ bounds according to WMAP for the relic density:
$0.0945<\Omega_{CDM}h^2<0.1287$. As a first result, we found that for $A_0=0$, 
sgn$\mu=+$, small values of tan$\beta$ are not favored; only for 
tan$\beta\simeq50$ are both criteria significantly consistent.
 
\end{abstract}

\maketitle

\section{Introduction}

One of the most accepted candidates to be the major component of dark matter 
(DM) is the neutralino as an LSP (Lightest Supersymmetric Particle). 
Supersymmetric models with R-parity conservations predict this type of 
particles (for an excellent introduction to Supersymmetry see \cite{martin}).
This type of models have several parameters that can be constrained in its
values using observational constraints of the actual density of DM, according
with WMAP: 
$0.0945\le\Omega_{CDM}h^2\le0.1287$ \cite{belanger,WMAP}. In particular for
mSUGRA models this has been done using the standard approach \cite{belanger,constraint} 
which is based in the Boltzmann
equation considering that after the ``freeze-out'' era, neutralinos cease
to annihilate keeping its number constant. In such an approach, the relic
density of neutralinos is approximately: $\Omega_{\chi}\approx 
1/\langle\sigma v\rangle$, where $\langle\sigma v\rangle$ is the thermally
averaged cross section times the relative velocity of the LSP annihilation
pair. Within the mSUGRA model five parameters ($m_0$, $m_{1/2}$, $A_0$, 
tan$\beta$ and the sign of $\mu$) are needed to specify the supersymmetric 
spectrum of particles and the final relic density. We will use the numerical
code micrOMEGAs \cite{micro} to compute the relic density following the past
scheme which will be called the ``abundance criterion'' (AC). 

Just after ``freeze-out'', we can consider neutralinos then as forming a 
Maxwell-Boltzmann (MB) gas in thermal equilibrium with other components of
the primordial cosmic structures. In the present time, such a gas is
almost colisionless and either constitutes galactic halos and larger structures
or it is in the process of its formation. In this context, we can conceive
two equilibrium states for the neutralino gas, the decoupling (or 
``freeze-out'') epoch and its present state as a virialized system. Computing
the entropy per particle for each one of this states we can use an 
``entropy consistency'' criterion (EC) using theoretical and empirical 
estimates for this entropy to obtain the relic density of neutralinos 
($\Omega_{\chi}$).

Our objective is then to use AC and EC criteria, to obtain constraints
for the parameters of the mSUGRA model by demanding that both criteria must
be consistent within them and within the observational constraints required by 
WMAP. 

\section{Abundance criterion}

Relic abundance of some stable species $\chi$ is defined as 
$\Omega_{\chi}=\rho_{\chi}/\rho_{crit}$, where $\rho_{\chi}=m_{\chi}n_{\chi}$
is the relic's mass density ($n_{\chi}$ is the number density), $\rho_{crit}$
is the critical density of the Universe (see \cite{kamion} for a review
on the standard method to compute the relic density). The time evolution of
$n_{\chi}$ is given by the Boltzmann equation:
\begin{equation}\label{bolt}
\frac{dn_{\chi}}{dt}=-3Hn_{\chi}-\langle\sigma v\rangle(n_{\chi}^2
-(n_{\chi}^{eq})^2)
\end{equation}
where $H$ is the Hubble expansion rate, $\langle\sigma v\rangle$ is the 
thermally averaged cross section times the relative velocity of the LSP 
annihilation pair and $n_{\chi}^{eq}$ is the number density that species would
have in thermal equilibrium. In the early Universe, the neutralinos ($\chi$) 
were initially in thermal equilibrium, $n_{\chi}=n_{\chi}^{eq}$. As the 
Universe expanded, their typical interaction rate started to diminish an the
process of annihilation froze out. Since then, the number density of neutralinos has remained
basically constant.

There are several ways to solve equation (\ref{bolt}), one of the more used
is based on the ``freeze-out'' approximation (see for example \cite{gondolo}).
However in order to have more precision, we will use the exact solution to 
Boltzmann equation using the public numerical code micrOMEGAs 1.3.6 
\cite{micro} which calculates the relic density of the LSP in the Minimal 
Supersymmetric Standard Model (MSSM). We will take and mSUGRA model and its 
five parameters ($m_0$, $m_{1/2}$, $A_0$, tan$\beta$ and the sign of $\mu$) 
as input parameters for micrOMEGAs and use {\it Suspect} \cite{suspect}, 
which comes as an interface to micrOMEGAs, to calculate the supersymmetric 
spectrum of masses of particles. Details about how we used micrOMEGAs
for making the calculation will be described in a future paper that is 
currently in preparation \cite{nuevo}. 

Using micrOMEGAs, we can obtain the relic density for any region of the 
parameter space to discriminate regions that are consistent with the WMAP 
constraints in this abundance criterion. 

\section{Entropy consistency criterion}

Since the usual MB statistics that can be formally applied to the neutralino gas
at the ``freeze-out'' era can not be used to describe present day neutralinos subject
to a long range gravitational interaction making up non-extensive systems, it is necessary
to use the appropriate approach that follows from the microcanonical ensemble in the 
``mean field'' approximation which yields an entropy definition that is well defined for a 
self-gravitating gas in an intermediate state. Such an approach is valid at both the initial 
(``freeze-out'' era, $f$) and final (virialized halo structures, $h$) states
that we wish to compare. Under these conditions, the change in the entropy per particle ($s$) between 
these two states is given by \cite{lyr}:
\begin{equation}\label{entropy}
s^h-s^f=ln\left[\frac{n_{\chi}^f}{n_{\chi}^h}\left(\frac{x^f}{x^h}\right)^{3/2}\right]
\end{equation}
where $x=m_{\chi}/T$, $T$ is the temperature of the gas. A region that fits with the conditions associated
with the intermediate scale is the central region of halos ($~10 pc^3$ within the halo core); evaluating
the thermodynamical quantities at this region, using equation (\ref{entropy}) and some 
assumptions more, it is possible to construct a theoretical estimate for $s^h$ that depends on the nature 
of neutralinos ($m_{\chi}$ and $\langle\sigma v\rangle$), initial conditions (given by $x^f$), cosmological 
parameters ($\Omega_{\chi}$, the Hubble parameter, $h$) and structural parameters of the virialized halo 
(central values for temperature and density); for details of these and the following, see section IV of 
\cite{lyr}. 

An alternative estimate for $s^h$ can be made based on empirical quantities for observed
structures in the present Universe using the microcanonical entropy definition in terms of phase space
volume, but restricting this volume to the actual range of velocities accessible to the central particles, that
is, up to a maximal escape velocity $v_e(0)$ which is related to the central velocity dispersion of the halo 
($\sigma_h$) by an intrinsic parameter $\alpha$: $v_e^2(0)\sim\alpha\sigma_h^2(0)$. The authors in \cite{lyr}
give an uncertainty range for the value of $\alpha$ for actual galaxies: $11.2\le\alpha\le24.8$. The range of
values allowed for this parameter is of the highest importance to determine the allowed region of the parameter
space in the mSUGRA model as will be clear in the results presented on next section. 

Equating the theoretical an empirical estimates for the entropy per particle it is obtained a relation for the relic abundance of neutralinos using the EC criterion\footnote{This formula is a small modification to the one 
presented in \cite{lyr}}:
\begin{equation}\label{consistency}
ln(\Omega_{\chi}h^2)=10.853-x^f+ln\left[\frac{(x^f\alpha)^{3/2}m_{\chi}}{f_g^*(x^f)}\right]
\end{equation}
where $f_g^*(x^f)$ is a function related to the degrees of freedom at the ``freeze-out'' time (see for example \cite{gondolo}) that will be described elsewhere \cite{nuevo}.

Modifying the program micrOMEGAs, we can obtain the value for $x^f$ for any region of the 
parameter space and then $\Omega_{\chi}$ using (\ref{consistency}), therefore we will be able to  
discriminate regions that are consistent with the WMAP constraints for the EC criterion.

\section{Results and Conclusions}

\begin{figure}\centering
\includegraphics[height=6cm, width=15cm]{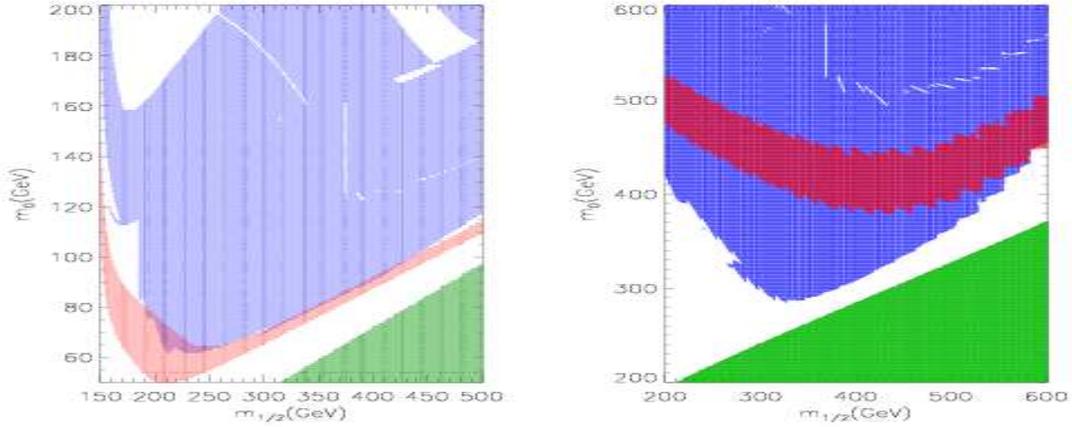}
\caption{Allowed regions in the parameter space for AC (red) and EC (blue) criteria for the 
mSUGRA model with $A_0=0$ and sgn$\mu=+$. The left panel shows the results for 
tan$\beta=10$ and the right one for tan$\beta=50$.}
\label{results}
\end{figure}

Using both AC and EC criteria, that have been briefly described above, we can compute the 
relic abundance of neutralinos and constrain the region in the mSUGRA parameter space where 
both criteria are fullfilled. In order to obtain a first result we will fix the value of two
parameters in the mSUGRA model: $A_0=0$ and sgn$\mu=+$. In the left panel of figure 
(\ref{results}), we present a region of the parameter space with these two values fixed and 
with tan$\beta=10$. The green region is where the $\tilde{\tau}$ is the 
LSP, the red and blue areas determine the WMAP allowed regions for the AC and EC 
criteria respectively. As we can see from figure (\ref{results}) (left panel), the region 
where both criteria are fullfilled is very small, in fact only for the highest values of 
$\alpha$ there is an intersection between both criteria. The right panel of figure 
(\ref{results}) shows the same regions as in the left panel but
for tan$\beta=50$; contrary to the case of tan$\beta=10$ we see now a complete intersection 
of both AC and EC criteria.

We have followed the novel idea of \cite{lyr} to introduce a new criterion to constrain the mSUGRA parameter space using the assumption of entropy consistency for the initial and final states of a neutralino gas. Using the program micrOMEGAs, we explored with precision which regions then satisfy this criterion and the usual AC criteria previously used several times. We found that for the regions so far explored, values with small tan$\beta$ are not favored, leading to an insignificant allowed region satisfying both criteria. Values with tan$\beta\gtrsim50$ fullfill the requirement of both criteria and the WMAP constraints. Further analysis, which is currently being done, is required to give more precise conclusions about this new method to constrain the parameter space of the mSUGRA model. 

We acknowledge partial support by CONACyT M\'exico, under grants 32138-E, 
34407-E and 42026-F, and PAPIIT-UNAM IN-122002, IN117803 and IN116202 grants. 
JZ acknowledges support from DGEP-UNAM and CONACyT scholarships.

\end{document}